# Open-Set Cross-Network Node Classification via Unknown-Excluded Adversarial Graph Domain Alignment

Xiao Shen[1], Zhihao Chen[1], Shirui Pan[2],
Shuang Zhou[3], Laurence T. Yang[4, 5], and Xi Zhou[1*]

[1]Hainan University
[2]Griffith University
[3]The Hong Kong Polytechnic University
[4]Zhengzhou University
[5]St. Francis Xavier University

xshen@hainanu.edu.cn, zhchen@hainanu.edu.cn, s.pan@griffith.edu.au,
shuang.zhou@connect.polyu.hk, ltyang@ieee.org, xzhou@hainanu.edu.cn

## Abstract

Existing cross-network node classification methods are mainly proposed for closed-set setting, where the source network and the target network share exactly the same label space. Such a setting is restricted in real-world applications, since the target network might contain additional classes that are not present in the source. In this work, we study a more realistic open-set cross-network node classification (O-CNNC) problem, where the target network contains all the known classes in the source and further contains several target-private classes unseen in the source. Borrowing the concept from open-set domain adaptation, all target-private classes are defined as an additional "unknown" class. To address the challenging O-CNNC problem, we propose an unknown-excluded adversarial graph domain alignment (UAGA) model with a separate-adapt training strategy. Firstly, UAGA roughly separates known classes from unknown class, by training a graph neural network encoder and a neighborhood-aggregation node classifier in an adversarial framework. Then, unknown-excluded adversarial domain alignment is customized to align only target nodes from known classes with the source, while pushing target nodes from unknown class far away from the source, by assigning positive and negative domain adaptation coefficient to known class nodes and unknown class nodes. Extensive experiments on real-world datasets demonstrate significant outperformance of the proposed UAGA over state-of-the-art methods on O-CNNC.

**Code** —https://github.com/3480430977/UAGA

## Introduction

Cross-network node classification (CNNC) (Shen et al. 2021) aims to leverage the knowledge obtained from a source network with rich labels to facilitate the classification of nodes in a target network that lacks labels. Existing CNNC methods (Shen et al. 2020; Zhang et al. 2021; Huang, Xu, and Wang 2022; Dai et al. 2023; Qiao et al. 2023; Shao et al. 2024; Shen et al. 2024) are mainly under the closed-set assumption that the source and target networks share an identical group of node classes. However, it is hard to guarantee such a closed-set setting in real-world CNNC applications, since we cannot determine whether the source and target networks share the same label space if no target labels are available.

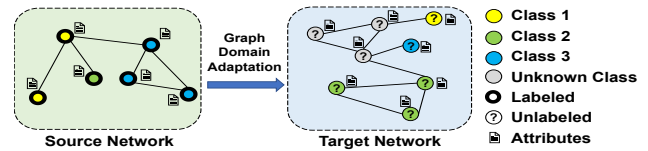

Figure 1: An illustration of the O-CNNC problem.

To break through the restriction of the closed-set assumption, this work studies a more realistic open-set cross-network node classification (O-CNNC) problem, where the target network contains not only all the known classes in the source network, but also an additional "unknown" class (Saito et al. 2018) covering all the target-private classes unseen in the source. Figure 1 illustrates the O-CNNC problem, the goal is to classify target nodes from known classes into the corresponding seen classes, and recognize target nodes belonging to target-private classes as "unknown".

To effectively address the O-CNNC problem, one need to solve two challenges in essence: 1) Since the target network is without any labels, we do not know which target nodes belong to unknown class. Thus, *how to generate a*

---

*boundary between known classes and unknown class is the first obstacle in O-CNNC.* 2) The domain discrepancy would impede a model trained on the source network to be directly applied to a new target network. Moreover, due to the absence of unknown class in the source, directly matching the whole distribution between the source and target networks like existing closed-set CNNC literature would be risky, since the misalignment between target unknown class and source known classes could cause negative transfer and make it more difficult to detect unknown class. Thus, *how to align the target distribution with the source while excluding unknown class is the key to success in O-CNNC.*

We propose a novel framework, named <u>u</u>nknown-excluded <u>a</u>dversarial <u>g</u>raph domain <u>a</u>lignment (UAGA), to address the challenging O-CNNC problem. The proposed UAGA is trained with a separate-adapt strategy, which roughly separates unknown class prior to unknown-excluded adversarial domain alignment. In separation stage, UAGA learns a rough boundary between unknown class and known classes, by training an attention-based graph neural network (GNN) encoder and a ($K$+1)-class neighborhood-aggregation node classifier in an adversarial framework. In adaptation stage, pseudo-label would be assigned to a target node, only if its clustering and classification prediction reach an agreement. Such refined target pseudo-labels would be taken as the supervised signals to iteratively retrain the model in a self-training manner, which yields a refined boundary to separate unknown class from known classes progressively. Moreover, instead of matching the whole distribution between the source and target networks, we propose to conduct unknown-excluded adversarial domain alignment to explicitly exclude unknown class from cross-network distribution matching. The conventional adversarial domain adaptation method (Ganin et al. 2016) always assigns positive domain adaptation coefficient to all samples across domains in the gradient reversal layer (GRL). In contrast, we innovatively propose to assign negative (positive) domain adaptation coefficient to nodes belonging to unknown (known) classes. On one hand, the positive domain adaptation coefficient guides the GNN encoder and domain discriminator to compete against each other, so as to learn network-invariant embeddings for known classes. On the other hand, the negative domain adaptation coefficient guides the GNN encoder and domain discriminator to be trained in the same direction to make the embeddings of target unknown class very distinguishable from the source embeddings. As a result, the proposed UAGA aligns the target distribution to the source only for known classes while excluding unknown class. The contributions of this work are summarized as follows:

**Problem.** We study a novel O-CNNC problem, and provide theoretical analysis of homophily and open-set domain adaptation (OSDA) w.r.t. O-CNNC.

**Algorithm.** We propose a novel separate-adapt framework named UAGA to address O-CNNC, which initially learns a rough boundary to separate unknown from known by adversarial learning, followed by unknown-excluded adversarial domain alignment. To our knowledge, UAGA is the first work to employ negative domain adaptation coefficient to exclude unknown class samples from adversarial domain adaptation. Empowered by such design, UAGA can reduce domain discrepancy only for known classes, while pushing target unknown class far away from the source to avoid negative transfer.

**Experiment.** Extensive experimental results demonstrate the superiority of the proposed UAGA over state-of-the-art baselines by a large margin.

## Related Work

**Open-set Node Classification** rejects unlabeled nodes not belonging to any known classes as an "unknown" class. OpenWGL (Wu, Pan, and Zhu 2020) performs uncertainty-aware node representation learning and separates unknown samples from known ones via a threshold based on the model's confidence scores. OODGAT (Song and Wang 2022) introduces an attention mechanism to explicitly distinguish inliers from outliers during feature propagation process of GNN. G$^2$Pxy (Zhang et al. 2023) follows an inductive learning setting where the information about unknown class is unavailable during model training. These open-set node classification methods are all designed for a single-network scenario. In contrast, our work studies open-set node classification in a cross-network scenario, where the source and target networks inherently exhibit diverse data distributions. Without addressing the domain discrepancy for known classes across networks, the single-network-based open-set node classification methods might exhibit sub-optimal performance in the challenging O-CNNC task.

**Open-set Domain Adaptation** allows target domain to contain new classes not present in the source. OSBP (Saito et al. 2018) is the most representative OSDA method which trains a classifier and a feature generator by adversarial learning, where the classifier is trained to construct a boundary between known class and unknown class samples, while the generator is trained to make target samples far away from the boundary. STA (Liu et al. 2019) progressively separates the samples of unknown class and known classes, and conducts weighted domain adaptation to only match the distribution of known classes across domains. OMEGA (Ru et al. 2023) designs unknown-aware target clustering to form tight clusters in target domain and generates specific thresholds for each target sample by moving-threshold estimation. Existing OSDA methods are mostly developed for computer vision (CV) field with the assumption of independent and identically distributed (i.i.d.) samples within each domain.

However, the graph-structured data obviously violate the i.i.d. assumption, due to complex network connections between nodes (Shen et al. 2020; Wu et al. 2020; Dai et al. 2023). Thus, directly applying existing OSDA methods to address the O-CNNC problem might fail to obtain satisfactory performance.

**Cross-network Node Classification**. CDNE (Shen et al. 2021) is the pioneering CNNC method which employs stacked auto-encoders to reconstruct the proximity matrix of each network. ACDNE (Shen et al. 2020) leverages dual feature extractors to separately learn self-representations from neighbor-representations. AdaGCN (Dai et al. 2023) integrates graph convolution network (GCN) with Wasserstein distance guided adversarial domain adaptation. UDAGCN (Wu et al. 2020) adopts dual GCNs to capture local and global consistency. DMGNN (Shen et al. 2023) devises label-aware propagation scheme to promote intra-class propagation while avoiding inter-class propagation. DGDA (Cai et al. 2024) applies variational graph auto-encoders to disentangle the semantic latent variables, domain latent variables, and random latent variables. A2GNN (Liu et al. 2024) removes propagation layers in source graph and stacks multiple propagation layers in target graph. The aforementioned CNNC methods are all designed upon the closed-set setting, i.e., the source and target networks share exactly the same label space. To reduce domain discrepancy, they adopt either statistical matching (Shen et al. 2021; Wu, He, and Ainsworth 2023) or adversarial learning (Shen et al. 2020; Wu et al. 2020; Dai et al. 2023) techniques to align the whole distribution of the target network with the source network. However, such whole distribution matching is not supposed to perform well in O-CNNC, since aligning target unknown samples with the source would cause negative transfer (Saito et al. 2018).

Very recently, some work investigate the CNNC problem without the closed-set restriction. UDANE (Chen et al. 2023) applies an entropy regularization on target nodes to enforce a separation between unknown class and known classes. SDA (Wang et al. 2024) separates target unknown class from known classes by neighbor center clustering.

## Preliminaries

Let $\mathcal{G}^s = (\mathcal{V}^s, \mathcal{E}^s, \mathbf{A}^s, \mathbf{X}^s, \mathbf{Y}^s)$ denote a fully labeled source network with a set of nodes $\mathcal{V}^s$ and a set of edges $\mathcal{E}^s$, where $\mathbf{A}^s \in \{0,1\}^{n^s \times n^s}$, $\mathbf{X}^s \in \mathbb{R}^{n^s \times \omega}$ and $\mathbf{Y}^s \in \{0,1\}^{n^s \times K}$ are the adjacency matrix, node attribute matrix and node label matrix of $\mathcal{G}^s$, $n^s$ and $\omega$ are the number of nodes and attributes, and $K$ is the number of known classes in $\mathcal{G}^s$. Let $\mathcal{G}^t = (\mathcal{V}^t, \mathcal{E}^t, \mathbf{A}^t, \mathbf{X}^t)$ denote an unlabeled target network with a node set $\mathcal{V}^t$ and an edge set $\mathcal{E}^t$, where $\mathbf{A}^t \in \{0,1\}^{n^t \times n^t}$ and $\mathbf{X}^t \in \mathbb{R}^{n^t \times \omega}$ are the adjacency matrix and node attribute matrix of $\mathcal{G}^t$, and $n^t$ is the number of nodes in $\mathcal{G}^t$.

**Definition 1. Open-set Cross-network Node Classification (O-CNNC).** Given a fully labeled $\mathcal{G}^s$ and a completely unlabeled $\mathcal{G}^t$ following two distinct distributions $\mathbb{P}^s$ and $\mathbb{Q}^t$. Let $\mathcal{Y}^s$ and $\mathcal{Y}^{to}$ denote the original label space of $\mathcal{G}^s$ and $\mathcal{G}^t$, where $\mathcal{Y}^{to} = \mathcal{Y}^s \cup \mathcal{Y}^u$ contains 1) all the known classes in $\mathcal{Y}^s = \{1, \cdots, K\}, K \geq 2$, and 2) some extra new classes $\mathcal{Y}^u$ ($|\mathcal{Y}^u| \geq 1$) unseen in $\mathcal{Y}^s$. Such target-private classes $\mathcal{Y}^u$ are all represented by the $(K+1)$-th "unknown" class, since we know nothing about these classes. Accordingly, one can obtain a new label space $\mathcal{Y}^t = \{1, \cdots, K, K+1\}$. The goal of O-CNNC is to learn an optimal classifier $\hbar: \mathcal{G}^t \to \mathcal{Y}^t$ such that 1) the target nodes whose labels belonging to $\mathcal{Y}^s$ are classified into one of the first $K$ known classes, and 2) the target nodes whose labels belonging to $\mathcal{Y}^u$ are recognized as the $(K+1)$-th "unknown" class.

**Definition 2. Homophily Ratio** (Pei et al. 2020). Node homophily ratio $\mathbb{h}$ is the average fraction of neighbors having the same class-label with each central node in graph $\mathcal{G}$:

$$\mathbb{h} = \frac{1}{|\mathcal{V}|} \sum_{v_i \in \mathcal{V}} \frac{\{j | j \in \mathcal{N}_i \wedge y_i = y_j\}}{|\mathcal{N}_i|} \quad (1)$$

where $\mathcal{N}_i = \{j | j \neq i, A_{i,j} = 1\}$ is a set of first-order neighbors of $v_i$, $y_i$ and $y_j$ are the class-label of $v_i$ and $v_j$. A large ratio $\mathbb{h}$ implies that $\mathcal{G}$ is more homophilic w.r.t. the node labels, i.e., connected nodes more tend to share the same class-label (McPherson, Smith-Lovin, and Cook 2001).

**Theorem 1. Homophilic w.r.t. $K+1$ Classes in O-CNNC**. Given a target graph $\mathcal{G}^t$ with original label space $\mathcal{Y}^{to} = \mathcal{Y}^s \cup \mathcal{Y}^u$, $\mathcal{Y}^s = \{1, \cdots, K\}$. Assume that there exists a mapping $\mathcal{f}: \mathcal{Y}^{to} \to \mathcal{Y}^t = \{1, \cdots, K, K+1\}$, where all classes in $\mathcal{Y}^u$ are represented by a new class $K+1$. If $\mathcal{G}^t$ is homophilic w.r.t. $\mathcal{Y}^{to}$, then $\mathcal{G}^t$ is also homophilic w.r.t. $\mathcal{Y}^t$.

**Proof**: According to Definition 2, the homophily ratio of $\mathcal{G}^t$ w.r.t. its original label space $\mathcal{Y}^{to}$ is measured as:

$$\mathbb{h}^{\mathcal{Y}^{to}} = \frac{1}{|\mathcal{V}^t|} \sum_{v_i \in \mathcal{V}^t} \frac{|\mathcal{N}_i^1|}{|\mathcal{N}_i|} \quad (2)$$

where $\mathcal{N}_i$ can be divided into two disjoint subsets: 1) $\mathcal{N}_i^1 = \{j | j \in \mathcal{N}_i \wedge y_i^{to} = y_j^{to}\}$ is a set of intra-class neighbors sharing the same class-label with $v_i$, 2) $\mathcal{N}_i^2 = \{j | j \in \mathcal{N}_i \wedge y_i^{to} \neq y_j^{to}\}$ is a set of inter-class neighbors having different class-labels with $v_i$, and $y_i^{to}, y_j^{to} \in \mathcal{Y}^{to}$ are the original labels of $v_i$ and $v_j$.

Let $\mathcal{f}(y_i^{to}), \mathcal{f}(y_j^{to}) \in \{1, \cdots, K, K+1\}$ denote new labels of $v_i$ and $v_j$ after mapping $\mathcal{f}: \mathcal{Y}^{to} \to \mathcal{Y}^t$. The homophily ratio of $\mathcal{G}^t$ w.r.t. new label space $\mathcal{Y}^t$ can be derived as:

$$\mathbb{h}^{\mathcal{Y}^t} = \frac{1}{|\mathcal{V}^t|} \sum_{v_i \in \mathcal{V}^t} \frac{|\mathcal{N}_i^1| + |\{j | j \in \mathcal{N}_i^2 \wedge \mathcal{f}(y_i^{to}) = \mathcal{f}(y_j^{to})\}|}{|\mathcal{N}_i|} \quad (3)$$

Since we have $|\{j | j \in \mathcal{N}_i^2 \wedge \mathcal{f}(y_i^{to}) = \mathcal{f}(y_j^{to})\}| \geq 0$ for any $v_j$, we can derive that $\mathbb{h}^{\mathcal{Y}^t} \geq \mathbb{h}^{\mathcal{Y}^{to}}$. Therefore, if $\mathcal{G}^t$ is homophilic w.r.t. original label space $\mathcal{Y}^{to}$, then $\mathcal{G}^t$ should be no less homophilic w.r.t. new label space $\mathcal{Y}^t$ containing $K$ known classes and the $(K+1)$-th "unknown" class.

*Remark:* According to Theorem 1, for target graph with homophily in O-CNNC, nodes no matter belonging to known classes or "unknown" class, tend to connect with others sharing the same class-label. Such connection patterns in graph have been shown to be informative for both known classification and unknown detection (Song and Wang 2022; Wu et al. 2023; Zhou et al. 2023). This naturally motivates us to adopt a GNN encoder and a ($K$+1)-class node classifier to aggregate neighborhood information for joint known classification and unknown detection in O-CNNC.

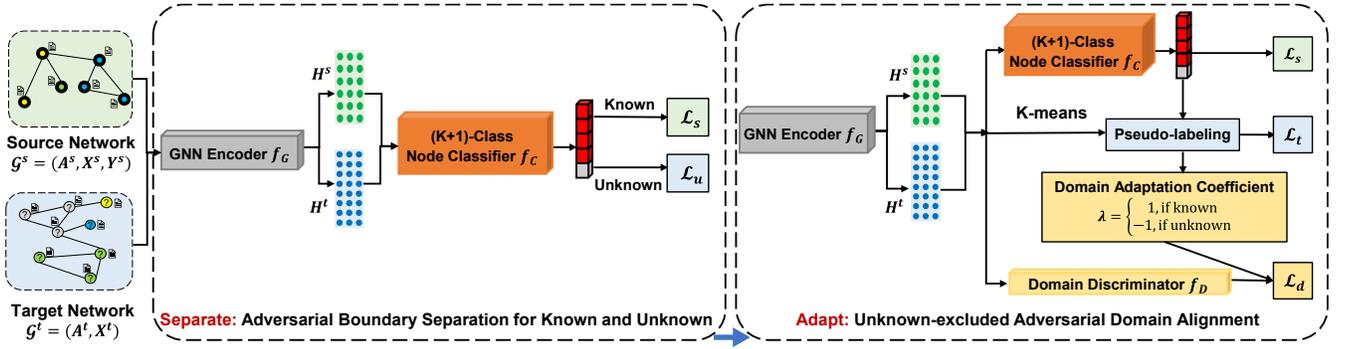

Figure 2: Model architecture of UAGA with a separate-adapt framework. In separation stage, a GNN encoder and a node classifier are trained in an adversarial manner to roughly separate unknown from known. In adaptation stage, positive and negative domain adaptation coefficients are leveraged to explicitly exclude unknown class from adversarial domain alignment.

**Theorem 2. OSDA Generalization Bounds** (Fang et al. 2020). Let $\mathbb{P}^s$ and $\mathbb{Q}^t$ denote the distribution of the source and target domains respectively, $\pi_{K+1}^t$ denote the class-prior probability of target unknown class. Given a hypothesis space $\mathcal{H}$ with a mild condition that the constant value function $K+1 \in \mathcal{H}$, for $\forall h \in \mathcal{H}$, the target risk $R_t(h)$ is bounded by:

$$\frac{R_t(h)}{1-\pi_{K+1}^t} \leq R_s(h) + disc(\mathbb{Q}_{X|Y\leq K}^t, \mathbb{P}_X^s) + \frac{\pi_{K+1}^t}{1-\pi_{K+1}^t} R_{t,K+1}(h) + \Lambda \quad (4)$$

where $R_s(h)$ is the source risk, $disc(\mathbb{Q}_{X|Y\leq K}^t, \mathbb{P}_X^s)$ is the domain discrepancy upon the $K$ known classes shared by the source and target domains, $(\pi_{K+1}^t/(1-\pi_{K+1}^t))R_{t,K+1}(h)$ is the open-set risk in target domain that can be interpreted as the mis-classification rate of unknown samples, and $\Lambda$ is the shared error of the joint optimal hypothesis for the source and target domains.

*Remark:* Note that O-CNNC can be seen as applying OSDA to node classification task across graphs $\mathcal{G}^s$ and $\mathcal{G}^t$. According to Theorem 2, in order to reduce target risk in O-CNNC, one should 1) minimize the classification error on $\mathcal{G}^s$, 2) match the domain distribution between $\mathcal{G}^s$ and $\mathcal{G}^t$ only for known classes, and 3) reduce the open-set risk on $\mathcal{G}^t$ (i.e., accurately identify "unknown" class nodes).

## The Proposed UAGA Model

In this section, we go through the details of the proposed UAGA to address O-CNNC. The framework of UAGA with a separate-adapt strategy is illustrated in Figure 2.

### Adversarial Boundary Separation for Known and Unknown

In separation stage, a GNN encoder $f_G$ and a ($K$+1)-class node classifier $f_C$ are trained in an adversarial manner to learn a rough boundary between unknown class and known classes. The proposed UAGA resorts to graph attention network (GAT) (Veličković et al. 2018) to construct the GNN encoder $f_G$ for node embedding learning, as:

$$\boldsymbol{h}_i = \text{ReLU}\left(\sum_{j \in \mathcal{N}_i \cup \{i\}} \tilde{A}_{i,j} \boldsymbol{W}^G \boldsymbol{x}_j\right) \quad (5)$$

where $\tilde{A}_{i,j}$ is adaptive edge weight of $(v_i, v_j)$ learned by GAT, $\boldsymbol{x}_i \in \mathbb{R}^\omega$ is input node attribute vector of $v_i$, $\boldsymbol{W}^G$ is learnable parameter, and $\boldsymbol{h}_i$ is node embedding vector of $v_i$.

The conventional OSDA methods (Saito et al. 2018; Feng et al. 2019; Shermin et al. 2020) typically construct a ($K$+1)-class task classifier by a multi-layer perceptron (MLP), to jointly deal with known classification and unknown detection. However, adopting an MLP-based classifier in O-CNNC fails to capture the homophily between connected nodes for graph data, since MLP considers each sample independently. According to Theorem 1, in O-CNNC, for a graph homophilic w.r.t. its original label space, it should be also homophilic w.r.t. the new label space containing $K$ known classes and ($K+1$)-th "unknown" class. Motivated by this, the proposed UAGA constructs a ($K$+1)-class neighborhood-aggregation node classifier $f_C$ by a graph attention layer (Veličković et al. 2018) to adaptively aggregate the logits among the neighborhood for all $K$+1 dimensions, as:

$$\hat{\boldsymbol{y}}_i = \text{Softmax}\left(\sum_{j \in \mathcal{N}_i \cup \{i\}} \tilde{A}_{i,j} \boldsymbol{W}^c \boldsymbol{h}_j\right) \quad (6)$$

where $\hat{\boldsymbol{y}}_i \in \mathbb{R}^{K+1}$ is the predicted label probability vector of $v_i$ after neighborhood aggregation, and $\boldsymbol{W}^c$ is learnable parameter. The source node classification loss is defined as:

$$\mathcal{L}_s = -\frac{1}{n^s} \sum_{i=1}^{n^s} \sum_{k=1}^{K} Y_{i,k}^s \log \hat{Y}_{i,k}^s \quad (7)$$

where $Y_{i,k}^s = 1$ if the ground-truth class-label of $v_i^s$ is $k$; otherwise, $Y_{i,k}^s = 0$. $\hat{Y}_{i,k}^s$ is the probability of $v_i^s$ belonging to class $k$ predicted by $f_C$ in Eq. (6). Minimizing $\mathcal{L}_s$ guides label-discriminative node embeddings to separate different known classes, consequently reducing the source risk.

**Adversarial Training of GNN Encoder and Node Classifier.** According to Theorem 2, to reduce the open-set risk on $\mathcal{G}^t$ which also partly bounds the target risk of O-CNNC, we need to recognize target nodes belonging to "unknown" class. To this end, we follow the representative OSDA method OSBP (Saito et al. 2018) to construct a rough boundary between known classes and unknown class by adversarial learning. On one hand, $f_C$ is trained to output unknown probability $\mu$ for each target node $v_i^t$, i.e., $\hat{Y}_{i,K+1}^t = \mu$, where $0 < \mu < 1$. On the other hand, $f_G$ is trained to maximize the error of $f_C$ by making $\hat{Y}_{i,K+1}^t$ very different from $\mu$, via two options: 1) increasing $\hat{Y}_{i,K+1}^t$ to be much larger than $\mu$ and then reject $v_i^t$ as unknown; or 2) decreasing $\hat{Y}_{i,K+1}^t$ to be much smaller than $\mu$ so as to classify $v_i^t$ into one of known classes. A binary cross-entropy loss is adopted to define the unknown classification loss, as:

$$\mathcal{L}_u = -\tfrac{1}{n^t}\sum_{i=1}^{n^t} \mu \log \hat{Y}_{i,K+1}^t + (1-\mu)\log(1-\hat{Y}_{i,K+1}^t) \quad (8)$$

The adversarial training of $f_G$ and $f_C$ can be achieved via optimizing the following objectives:

$$\min_{\theta_C}\{\mathcal{L}_s + \mathcal{L}_u\}, \min_{\theta_G}\{\mathcal{L}_s - \mathcal{L}_u\} \quad (9)$$

where $\theta_G$ and $\theta_C$ are the learnable parameters of $f_G$ and $f_C$. In Eq. (9), $\theta_C$ is trained to minimize $\mathcal{L}_u$ so as to construct a rough boundary between known and unknown, whereas $\theta_G$ is trained to maximize $\mathcal{L}_u$ to push target nodes far away from boundary. To update $\theta_G$ and $\theta_C$ simultaneously, we follow (Saito et al. 2018) to insert a GRL (Ganin et al. 2016) to flip the sign of the gradient during back-propagation.

### Unknown-excluded Adversarial Domain Alignment

After rough separation, we go into adaptation stage to align the target distribution to the source only for known classes while explicitly excluding unknown class.

While in O-CNNC, the target network is completely unlabeled, we do not know which target nodes belong to unknown class. Thus, we propose to assign pseudo-labels to target nodes beforehand. Firstly, we employ K-means algorithm to cluster all target nodes into $K+1$ clusters, by taking the embeddings learned by $f_G$ as inputs. For the first $K$ clusters corresponding to known classes, the initial centroid of each $k$-th cluster is simply computed as the average over the source embeddings of class $k$, i.e., $\mathbb{C}_k = \sum_{i=1}^{n^s} Y_{i,k}^s \boldsymbol{h}_i^s / \sum_{i=1}^{n^s} Y_{i,k}^s$. However, the last $(K+1)$-th cluster corresponds to "unknown" class, which is not present in $\mathcal{G}^s$. Therefore, we pick out top $R$ target nodes with the highest predicted unknown probability to form a pseudo-unknown set $\mathcal{U} = \{v_i^t | \hat{Y}_{i,K+1}^t \text{ is top } R \text{ highest}\}$. Then, the average embedding of $\mathcal{U}$ is adopted to compute the initial centroid of the $(K+1)$-th cluster, i.e., $\mathbb{C}_{K+1} = \sum_{v_i^t \in \mathcal{U}} \boldsymbol{h}_i^t / R$.

Given $K+1$ initial cluster centroids, each target node would be assigned to its closest centroid, consequently obtaining a cluster-label matrix $\hat{Y}^{t(clu)} \in \{0,1\}^{n^t \times (K+1)}$. Raw cluster-labels might contain noise. To obtain more accurate pseudo-labels, we propose to only assign a pseudo-label to a target node $v_i^t$, if and only if its cluster-label and class-label predicted by $f_C$ reach an agreement:

$$\bar{\bar{Y}}_{i,k}^t = \begin{cases} 1, \text{if } \hat{Y}_{i,k}^{t(clu)} = 1 \wedge \underset{j}{\arg\max}\, \hat{Y}_{i,j}^t = k \\ 0, \text{ otherwise} \end{cases} \quad (10)$$

Besides, we further assign pseudo-unknown label to top $R$ target nodes in the pseudo-unknown set $\mathcal{U}$, if they have not been assigned with any pseudo-labels by Eq. (10). Such confident pseudo-labeled target nodes are employed to iteratively re-train $f_G$ and $f_C$ in a self-training manner (Chen, Weinberger, and Blitzer 2011; Shen, Mao, and Chung 2020), by minimizing the following target node classification loss:

$$\mathcal{L}_t = -\tfrac{1}{n_l^t}\sum_{i=1}^{n_l^t}\sum_{k=1}^{K+1} \bar{\bar{Y}}_{i,k}^t \log \hat{Y}_{i,k}^t \quad (11)$$

where $n_l^t$ is the number of target nodes assigned with pseudo-labels. Minimizing $\mathcal{L}_t$ exploits the supervision from $\mathcal{G}^t$ to yield a refined boundary to separate $K+1$ classes, which is conductive to reducing the open-set risk on $\mathcal{G}^t$.

According to Theorem 2, to succeed in O-CNNC, it is essential to match the domain distribution between $\mathcal{G}^s$ and $\mathcal{G}^t$ only for known classes. Note that the OSBP method (Saito et al. 2018) adopted in previous separation stage does not utilize any domain information during adversarial learning (Shermin et al. 2020), thus it cannot explicitly reduce the domain discrepancy for known classes. To remedy this, we further employ a domain discriminator $f_D$ to compete against $f_G$, following conventional closed-set CNNC methods (Shen et al. 2020; Wu et al. 2020). The domain discriminator $f_D$ is constructed by an MLP taking node embeddings learned by $f_G$ as inputs and outputs $\hat{d}_i = f_D(\boldsymbol{h}_i; \theta_D)$, where $\hat{d}_i$ is the predicted probability of $v_i$ coming from $\mathcal{G}^t$. The domain classification loss is defined as:

$$\mathcal{L}_d = -\tfrac{1}{n^s+n^t}\sum_{i=1}^{n^s+n^t} d_i \log \hat{d}_i + (1-d_i)\log(1-\hat{d}_i) \quad (12)$$

where $d_i = 0$ if $v_i \in \mathcal{V}^s$ and $d_i = 1$ if $v_i \in \mathcal{V}^t$.

Then, $f_D$ and $f_G$ are trained in an adversarial manner by optimizing the following minimax objective:

$$\min_{\theta_G, \theta_C}\left\{\mathcal{L}_s + \beta\mathcal{L}_t + \lambda \max_{\theta_D}\{-\mathcal{L}_d\}\right\} \quad (13)$$

where $\beta$ and $\lambda$ are trade-off parameters to balance the effects of different losses. To simultaneously update the learnable parameters of $f_G$ and $f_D$, a GRL (Ganin et al. 2016) is inserted between them during back-propagation to reverse the gradient of $\mathcal{L}_d$ w.r.t. $\theta_G$ and multiply it by a domain adaptation coefficient $\lambda$.

**Positive and Negative Domain Adaptation Coefficient.** It is worth noting that in the conventional GRL-based adversarial domain adaptation method (Ganin et al. 2016), the domain adaptation coefficient $\lambda$ is always set to a positive value for all samples across domains. We argue that such a setting should be problematic in O-CNNC, since setting a positive $\lambda$ for all samples would align the entire target network with the source network without excluding unknown class, consequently causing negative transfer. To remedy this, we propose to assign positive $\lambda$ to nodes belonging to known classes, while negative $\lambda$ to nodes coming from unknown class, as:

$$\lambda = \begin{cases} 1, \text{if } v_i \in \mathcal{V}^s \vee \left(v_i \in \mathcal{V}^t \wedge \bar{\bar{Y}}^t_{i,K+1} = 0\right) \\ -1, \text{if } v_i \in \mathcal{V}^t \wedge \bar{\bar{Y}}^t_{i,K+1} = 1 \end{cases} \quad (14)$$

For simplicity, here we just assign fixed 1/-1 as the positive/negative value of $\lambda$. It is flexible to assign adaptive positive/negative values to $\lambda$, which we leave as the future work. On one hand, by assigning positive $\lambda$ to the source nodes and the target nodes with pseudo-known labels, $f_G$ and $f_D$ would be trained to compete against each other to yield network-invariant embeddings for known classes, like conventional closed-set domain adaptation method (Ganin et al. 2016). On the other hand, assigning negative $\lambda$ to the target nodes with pseudo-unknown label would train both $f_G$ and $f_D$ in the same direction to minimize $\mathcal{L}_d$, thus making the embeddings of target unknown class very distinguishable from the source embeddings. As a result, UAGA can reduce the domain discrepancy only for the shared known classes, while pushing target unknown class far away from the source to effectively reduce the open-set risk on $\mathcal{G}^t$, which is conductive to reducing the target risk of O-CNNC, according to Theorem 2.

**Algorithm Description**. Algorithm 1 shows the training process of UAGA in a mini-batch strategy. In separation stage (Line 2-9), the GNN encoder $f_G$ and the $(K+1)$-class neighborhood aggregation node classifier $f_C$ are trained in an adversarial manner via optimizing Eq. (9), so as to learn a rough boundary between unknown and known classes. In adaptation stage (Line 10-22), the GNN encoder $f_G$, node classifier $f_C$ and domain discriminator $f_D$ are trained via optimizing the overall minimax objective in Eq. (13). The target nodes would be assigned with pseudo-labels, according to the prediction results of both unsupervised clustering and supervised classification (Line 11-14). On one hand, such pseudo-labeled target nodes are employed to compute the target node classification loss $\mathcal{L}_t$ to iteratively re-train $f_G$ and $f_C$. On the other hand, such pseudo-labels are leveraged to assign positive/negative domain adaptation coefficient $\lambda$ to known/unknown class nodes. Thanks to unknown-excluded adversarial domain alignment and iterative self-training, a refined node classification boundary can be progressively learned. After training convergence or reaching the maximum training epochs of adaptation stage, the optimized parameters of $f_G$ and $f_C$ would be employed to generate node embeddings and predict target labels (Line 23-24).

**Complexity Analysis of UAGA.** Both $f_G$ and $f_C$ are constructed by a single graph attention layer. The time complexity of $f_G$ is $O\big((|\mathcal{V}^s| + |\mathcal{V}^t|)\omega\mathbb{d} + (|\mathcal{E}^s| + |\mathcal{E}^t|)\mathbb{d}\big)$,

---

**Algorithm 1** UAGA

**Input**: Fully labeled source network $\mathcal{G}^s = (A^s, X^s, Y^s)$ and unlabeled target network $\mathcal{G}^t = (A^t, X^t)$.

1. Initialize parameters $\theta_G, \theta_C, \theta_D$.
2. **while** *not max epoch of separation stage* **do**
3.    **while** *not max iteration* **do**
4.      Sample minibatch of $\mathbb{B}$ source nodes from $\mathcal{G}^s$ and minibatch of $\mathbb{B}$ target nodes from $\mathcal{G}^t$.
5.      Learn embeddings by $f_G$ in Eq. (5).
6.      Calculate $\mathcal{L}_s$ and $\mathcal{L}_u$ in Eq. (7) and Eq. (8).
7.      Update $\theta_G$ and $\theta_C$ in Eq. (9).
8.    **end while**
9. **end while**
10. **while** *not max epoch of adaptation stage* **do**
11.    Compute initial centroids of first $K$ clusters based on source embeddings $\{\mathbb{C}_k\}_{k=1}^K$.
12.    Compute initial centroid of $(K+1)$-th cluster $\mathbb{C}_{K+1}$ based on average embeddings of pseudo-unknown set $\mathcal{U}$.
13.    Apply $K$-means to obtain target cluster-labels $\widehat{Y}^{t(clu)}$.
14.    Consider both cluster-labels and class-labels predicted by $f_C$ to obtain refined pseudo-labels $\bar{Y}^t$ in Eq. (10).
15.    Assign positive or negative $\lambda$ in Eq. (14).
16.    **while** *not max iteration* **do**
17.      Sample minibatch of $\mathbb{B}$ source nodes from $\mathcal{G}^s$ and minibatch of $\mathbb{B}$ target nodes from $\mathcal{G}^t$.
18.      Learn embeddings by $f_G$ in Eq. (5).
19.      Calculate $\mathcal{L}_s, \mathcal{L}_t$ and $\mathcal{L}_d$ in Eq. (7), Eq. (11) and Eq. (12).
20.      Update $\theta_G, \theta_C, \theta_D$ in Eq. (13).
21.    **end while**
22. **end while**
23. Apply optimized $\theta_G^*$ to generate embeddings of $\mathcal{G}^s$ and $\mathcal{G}^t$ in Eq. (5).
24. Apply optimized $\theta_C^*$ to predict target labels in Eq. (6).

**Output**: Predicted node labels of target network $\widehat{Y}^t$.

---

where $|\mathcal{V}^s|$ and $|\mathcal{V}^t|$ are the number of nodes in $\mathcal{G}^s$ and $\mathcal{G}^t$ respectively, $|\mathcal{E}^s|$ and $|\mathcal{E}^t|$ are the number of edges in $\mathcal{G}^s$ and $\mathcal{G}^t$ respectively, $\omega$ is the number of node attributes, and $\mathbb{d}$ is the number of embedding dimensions. The time complexity of $f_C$ is $O\big((|\mathcal{V}^s| + |\mathcal{V}^t|)\mathbb{d}(K+1) + (|\mathcal{E}^s| + |\mathcal{E}^t|)(K+1)\big)$, where $K$ is the number of known classes. The time complexity of K-means clustering is $O\big((K+1)\mathbb{d}|\mathcal{V}^t|\big)$. The time complexity of $f_D$ constructed by an MLP is linear to the number of nodes across networks. Hence, the overall time complexity of UAGA is linear to number of nodes and edges in $\mathcal{G}^s$ and $\mathcal{G}^t$.

# Experiments

**Datasets**. Previous benchmark datasets for closed-set CNNC (Shen et al. 2021) only contain 5 node classes, which limits possible splits of known and unknown classes (i.e. various openness) in the open-set setting. To remedy this, we construct new benchmark datasets to contain more node classes for O-CNNC, i.e., Citation-v1 (C), DBLP-v4 (D) and ACM-v8 (A). They are real-world paper citation networks extracted from ArnetMiner with the papers published in different periods, i.e., between years 1997 and 2003, between years 2004 and 2011, and between years 2012 and 2015, respectively. Each node represents a paper and each edge denotes the citation relation between two papers. There are no common nodes between any two networks. The citation networks were modeled as undirected networks in our experiments. The keywords extracted from the paper title were utilized as node attributes. Each paper belongs to one of the following nine categories according to its research topics, including "Artificial Intelligence", "Human-computer Interaction", "Information Security", "Data Mining", "Computer Architecture", "Multimedia", "Computer Theory", "Computer Network", and "Software Engineering". Six O-CNNC tasks can be conducted among three networks, by selecting one as the source and another as the target, i.e., C→D, C→A, D→C, D→A, A→C and A→D. For each O-CNNC task, we chose the first $K$ classes as known classes, while all the remaining $9-K$ classes were re-labeled as the $(K+1)$-th "unknown" class, following the common setting in OSDA (Liu et al. 2019). The openness (Liu et al. 2019) is defined as the proportion of target-private classes in all original target classes, i.e., $\mathcal{O} = |\mathcal{Y}^u|/|\mathcal{Y}^{to}| = (9-K)/9$.

**Baselines**. The proposed UAGA is competed against 9 baselines in four categories: 1) <u>Open-set Domain Adaptation</u>: **OSBP** (Saito et al. 2018) and **OMEGA** (Ru et al. 2023), 2) <u>Open-set Node Classification</u>: **OODGAT** (Song and Wang 2022) and **G²Pxy** (Zhang et al. 2023), 3) <u>Closed-set Cross-network Node Classification</u>: **UDAGCN** (Wu et al. 2020), **AdaGCN** (Dai et al. 2023) and **SGDA** (Qiao et al. 2023), 4) <u>Open-set Cross-network Node Classification</u>: **SDA** (Wang et al. 2024) and **UDANE** (Chen et al. 2023).

**Implementation Details**. All experiments were conducted on a single Tesla A40 GPU with 48GB memory. The proposed UAGA was implemented by PyTorch 1.7.1 (Paszke et al. 2019) and Deep Graph Library 0.7.2 (Wang et al. 2019). UAGA was trained by the Adam optimizer with learning rate of 1e-3. The batch size $\mathbb{B}$ was set to 2048. The number of training epochs of separation stage and adaptation stage were set as 30 and 200 respectively. The number of layers of the GNN encoder $f_G$ was set to 1. The number of attention heads in GNN encoder $f_G$ and node classifier $f_C$ were set to 8 and 2 respectively. The number of embedding dimensions of each head $\mathbb{d}$ in the GNN encoder $f_G$ was set to 32. The weight of target node classification loss $\mathcal{L}_t$ (i.e.

| Datasets | # Nodes | # Edges | # Attributes | # Labels |
|---|---|---|---|---|
| Citation-v1 | 9737 | 14054 | | |
| DBLP-v4 | 8653 | 12967 | 7786 | 9 |
| ACM-v8 | 8806 | 17661 | | |

Table 1. Statistics of the experimental datasets.

$\beta$) was set to 0.1. The unknown threshold $\mu$ in rough separation stage was set to 0.5 following OSBP (Saito et al. 2018). The number of top $R$ target nodes to form pseudo-unknown set $\mathcal{U}$ was selected from {2000, 3000, 4000}.

**Evaluation Metrics**. We adopt three widely used metrics in OSDA literature to evaluate the O-CNNC performance. OS* (Saito et al. 2018) is the accuracy averaged over all known classes. OS (Saito et al. 2018) is the average accuracy over all classes. HS (Fu et al. 2020) is the harmonic mean of the instance-level accuracy on known classes and unknown class. We also adopt the Area Under the Receiver Operating Characteristic curve (AUC) to evaluate the detection of unknown class, following (Song and Wang 2022; Zhou et al. 2022). For each O-CNNC task, we run each method with 5 random initializations and report the average results and standard deviations.

## O-CNNC Results

Table 2 shows O-CNNC results under the openness $\mathcal{O} = 4/9$, i.e., the number of target-private classes is 4 and the number of known classes is 5. We have key observations as follows:

1) The OSDA baselines, i.e., OSBP and OMEGA perform poorly in O-CNNC. This is because they are developed based on the i.i.d. assumption, which ignores the connections between nodes in graph. However, considering the connection patterns between nodes has been proved to be essential for both known classification and unknown detection on graph data (Song and Wang 2022).

2) The open-set node classification baselines, i.e., OODGAT and G²Pxy, are good at unknown detection (high AUC) while rather weak in known classification (low OS*) in O-CNNC. This is because they are developed for a single-network scenario, failing to mitigate the domain discrepancy for known classes across networks.

3) The closed-set CNNC baselines (i.e., UDAGCN, AdaGCN, and SGDA) are good at known classification (high OS*) while still weak in unknown detection (low AUC). We believe the reason behind is that the domain adaptation component in the closed-set CNNC methods tends to match the whole distribution between the source and target networks, that is, the unknown class from target network would also be aligned with the source, which inevitably causes negative transfer for unknown detection.

4) The state-of-the-art O-CNNC baselines (i.e. UDANE and SDA) significantly outperform other baselines. Moreover, the proposed UAGA beats UDANE and SDA by a large margin across six O-CNNC tasks. The reason behind our outperformance might be two-fold. On one hand, to learn node embeddings, UDANE and SDA adopt GCN and dual-GCN, while UAGA adopts GAT as the GNN encoder. The latest literature on open-set node classification (Huang, Wang, and Fang 2022; Song and Wang 2022) have revealed that the attention-based model like GAT can achieve better performance on unknown detection than GCN. This is because GCN treats all neighbors equally during neighborhood aggregation, while GAT adaptively assigns small attention weight to neighbors from different distributions. On the other hand, to determine whether a target node belongs to one of the known classes or the "unknown" class, UDANE and SDA employ a thresholding method based on a $K$-class classifier which makes prediction over known classes. However, finding an optimal threshold to separate unknown class from known classes is hard and time-consuming (Zhang et al. 2023). Instead, our UAGA constructs a $(K + 1)$-class classifier by adding an extra class to represent the "unknown" class, which eliminates the difficulty of tuning the threshold.

| Task | Metric | OSBP | OMEGA | OODGAT | G²Pxy | UDAGCN | AdaGCN | SGDA | UDANE | SDA | UAGA |
|---|---|---|---|---|---|---|---|---|---|---|---|
| C→D | OS | 44.91±0.42 | 37.96±1.80 | 36.49±0.39 | 37.88±8.70 | 55.08±7.33 | 51.17±1.63 | 49.57±5.28 | 57.09±0.36 | 54.38±0.94 | **69.21±0.17** |
| | OS* | 52.36±0.44 | 30.38±2.62 | 27.02±0.46 | 36.52±9.89 | 63.14±10.24 | 58.63±1.81 | 56.68±6.42 | 61.78±0.59 | 58.94±1.27 | **74.00±0.15** |
| | AUC | 48.78±1.07 | 60.18±1.24 | 70.36±0.44 | 58.33±1.45 | 54.01±9.98 | 54.87±1.49 | 52.46±3.87 | 55.90±0.46 | 61.45±2.30 | **73.72±1.03** |
| | HS | 13.59±0.56 | 47.06±1.55 | 51.73±0.90 | 36.40±17.96 | 21.97±12.23 | 22.78±2.71 | 22.74±1.82 | 43.85±0.78 | 41.95±6.02 | **56.41±0.49** |
| C→A | OS | 41.98±1.07 | 29.35±1.28 | 28.25±0.55 | 39.83±6.30 | 44.33±13.68 | 45.96±0.77 | 45.80±4.89 | 54.76±0.69 | 54.93±1.87 | **69.50±0.39** |
| | OS* | 45.33±0.46 | 25.07±2.47 | 18.76±0.80 | 40.02±7.60 | 48.41±22.00 | 51.13±1.53 | 52.32±6.26 | 58.03±0.85 | 59.35±2.89 | **72.48±0.47** |
| | AUC | 63.17±0.57 | 54.89±0.55 | 55.45±0.75 | 60.46±1.86 | 56.49±6.21 | 61.69±1.33 | 50.39±3.26 | 62.13±0.60 | 66.88±1.95 | **83.41±0.26** |
| | HS | 33.51±1.04 | 38.65±0.71 | 35.39±1.25 | 32.59±16.39 | 16.58±16.22 | 29.46±4.22 | 21.19±2.62 | 46.53±0.24 | 41.36±9.43 | **61.39±0.60** |
| D→C | OS | 46.04±0.43 | 38.17±0.91 | 39.07±0.36 | 44.86±0.81 | 46.31±14.87 | 55.42±0.64 | 50.74±4.15 | 63.65±0.25 | 58.65±0.86 | **69.12±0.36** |
| | OS* | 51.63±0.60 | 31.48±3.39 | 29.89±0.56 | 43.98±2.05 | 51.17±19.21 | 61.41±1.03 | 57.50±5.13 | 68.08±0.22 | 63.01±1.53 | **74.76±0.21** |
| | AUC | 52.83±1.27 | 62.28±3.29 | 70.50±0.12 | 62.02±0.33 | 61.00±4.41 | 57.66±2.16 | 57.96±2.47 | 66.84±0.57 | 69.58±0.58 | **72.61±0.90** |
| | HS | 27.02±1.61 | 43.77±1.43 | 48.51±0.37 | 48.36±2.43 | 24.44±12.39 | 36.00±2.78 | 26.02±6.06 | 50.89±0.60 | 46.97±3.41 | **51.69±1.03** |
| D→A | OS | 43.71±1.48 | 33.96±0.56 | 27.62±0.29 | 42.73±0.91 | 48.81±6.31 | 55.09±0.45 | 53.70±4.33 | 60.22±1.00 | 55.95±1.03 | **64.83±1.16** |
| | OS* | 49.16±2.97 | 27.39±1.73 | 15.74±0.36 | 41.13±1.79 | 57.74±7.25 | 60.10±0.69 | 63.02±4.24 | 63.41±0.96 | 60.41±1.20 | **69.17±0.82** |
| | AUC | 57.18±2.61 | 59.95±0.92 | 61.46±0.26 | 61.15±0.11 | 56.32±4.53 | 65.21±1.16 | 55.75±3.97 | 69.25±1.28 | 70.90±0.39 | **78.33±2.61** |
| | HS | 23.98±7.07 | 43.65±0.97 | 37.40±0.60 | 47.39±0.90 | 6.96±8.93 | 40.50±2.24 | 10.93±14.13 | 51.54±1.23 | 43.85±0.87 | **51.90±2.54** |
| A→C | OS | 46.20±0.93 | 34.41±0.48 | 31.43±0.51 | 44.92±4.29 | 50.08±14.52 | 59.39±0.54 | 53.32±1.99 | 65.75±0.30 | 60.52±0.92 | **71.78±0.42** |
| | OS* | 51.50±2.56 | 32.20±0.57 | 19.90±0.68 | 43.01±7.95 | 54.27±20.73 | 66.50±0.77 | 58.66±2.13 | 69.78±0.39 | 65.44±2.34 | **77.58±0.44** |
| | AUC | 52.15±1.62 | 56.35±2.19 | 65.32±0.27 | 60.80±0.45 | 57.94±6.23 | 59.40±0.88 | 52.73±4.28 | 66.37±0.84 | 65.95±1.40 | **76.26±0.62** |
| | HS | 27.02±7.83 | 39.75±1.20 | 30.56±0.83 | 45.88±4.11 | 28.73±14.17 | 34.97±1.64 | 35.61±5.53 | 54.03±0.85 | 45.42±4.76 | **54.20±0.40** |
| A→D | OS | 44.47±1.08 | 37.11±1.12 | 30.79±0.12 | 47.12±2.97 | 52.65±6.02 | 55.56±0.74 | 46.99±4.09 | 58.42±0.60 | 55.71±0.85 | **66.40±0.34** |
| | OS* | 49.80±1.65 | 34.67±3.84 | 18.61±0.13 | 47.70±6.01 | 60.02±8.86 | 62.48±1.15 | 51.02±4.78 | 62.36±0.38 | 60.14±2.10 | **70.95±0.38** |
| | AUC | 52.24±1.44 | 56.12±4.78 | 71.67±0.58 | 59.17±0.17 | 54.95±3.48 | 56.62±0.95 | 51.34±2.14 | 62.07±3.81 | 64.16±1.27 | **71.87±1.40** |
| | HS | 26.71±2.16 | 40.95±1.09 | 35.89±0.73 | 44.83±7.29 | 20.62±16.18 | 31.73±3.49 | 36.11±2.52 | 47.92±2.45 | 43.85±4.57 | **53.93±0.42** |

Table 2: O-CNNC results on six tasks when openness is $\mathcal{O}=4/9$. The best scores among all methods are shown in boldface.

| Model Variants | C→D | | | | C→A | | | |
|---|---|---|---|---|---|---|---|---|
| | OS | OS* | AUC | HS | OS | OS* | AUC | HS |
| UAGA | 69.2 | 74.0 | 73.7 | 56.4 | 69.5 | 72.5 | 83.4 | 61.4 |
| w/o separation | 64.3 | 68.9 | 69.2 | 52.2 | 46.8 | 50.8 | 64.9 | 36.4 |
| w/o adaptation | 55.0 | 60.5 | 67.9 | 39.1 | 50.5 | 52.7 | 75.2 | 47.1 |
| w/o $\lambda = -1$ | 66.7 | 74.1 | 72.5 | 42.8 | 64.4 | 72.6 | 80.9 | 35.1 |
| Replace $f_c$ by MLP | 59.6 | 71.5 | 63.2 | 0.03 | 67.8 | 70.8 | 81.2 | 58.9 |
| w/o self-training | 58.2 | 62.9 | 63.9 | 44.9 | 54.8 | 57.5 | 73.1 | 48.7 |

Table 3. Model variants of UAGA.

**Various Openness**. We investigate the O-CNNC performance under various openness. As shown in Figure 3, the increase of openness generally leads to an increase of OS* while a decrease of AUC. This is because a larger openness implies more target-private classes and less known classes, which naturally eases known classification while increasing the difficulty of unknown detection. Moreover, one can see that the proposed UAGA consistently achieves better overall performance than all baselines under various openness. This reflects that UAGA allows openness-robust O-CNNC for both known classification and unknown detection.

**Ablation Study**. We investigate the contributions of different components in the proposed UAGA and report the results in Table 3. Removing separation stage results in significantly worse performance. This is because without learning a rough separation boundary between unknown and known beforehand, it would fail to obtain pseudo-labels to guide the following adaptation stage. Without adaptation stage, the performance also drops considerably. This shows that only rough separation is not enough, while further conducting unknown-excluded adversarial domain alignment with self-training is indispensable for UAGA. Without setting $\lambda = -1$ in the GRL-based adversarial domain adaptation leads to worse performance. This confirms that assigning negative domain adaptation coefficient to target unknown-class nodes to explicitly exclude them from adversarial domain alignment is a crucial component of UAGA. Replacing the neighborhood-aggregation node classifier $f_C$ by an MLP presents inferior performance. This is consistent with Theorem 1, i.e., $\mathcal{G}^t$ is homophilic w.r.t. $K+1$ classes. Thus, aggregating label prediction among neighborhood for all $K+1$ dimensions can indeed boost both known classification and unknown detection. Without self-training significantly degrades the performance. This verifies that exploiting confident pseudo-labeled target nodes to iteratively re-train the model indeed yields a refined boundary.

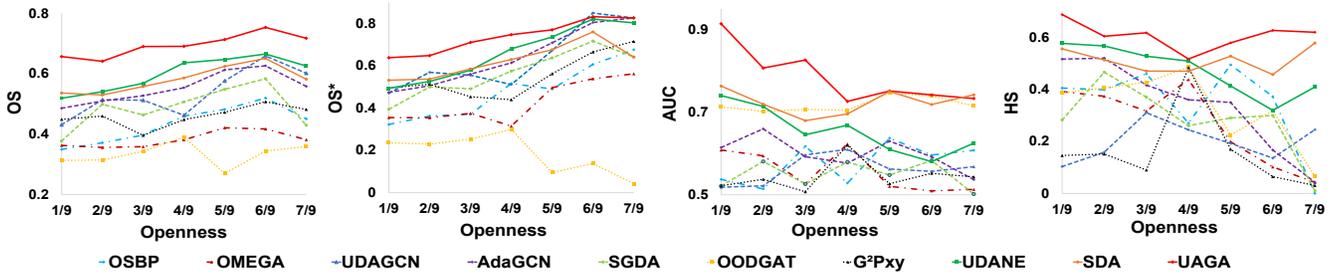

Figure 3: Performance of O-CNNC under various openness on the representative task D→C.

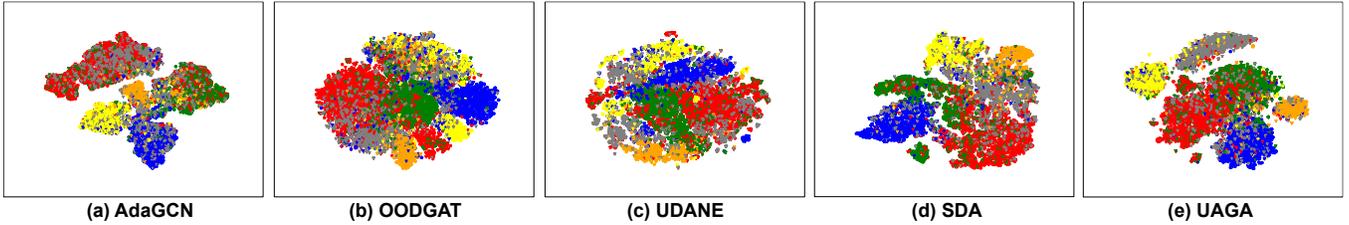

Figure 4: Visualization of cross-network embeddings on the representative task C→A. Grey indicates "unknown" class and other colors indicate different known classes.

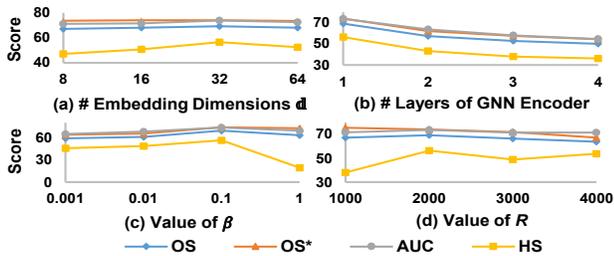

Figure 5. Parameter sensitivity of UAGA on task C→D.

**Visualization**. The cross-network node embeddings learned by representative methods were visualized by t-SNE (Van der Maaten and Hinton 2008) in Figure 4. We can observe that AdaGCN forms clear clusters between different known classes, but fails to detect unknown class. OODGAT fails to align the same known class of nodes across networks. UDANE and SDA can separate unknown class from known classes to some extent, however, the boundaries between different classes are not clear enough. The proposed UAGA yields the best visualization for both known classes and unknown class, where the target embeddings of known classes are aligned to the source accurately, while the embeddings of unknown class are separated far apart.

**Parameter Sensitivity**. As shown in Figure 5, UAGA achieves the best performance when the number of embedding dimensions $\mathbb{d}$ is 32. Deeper GNN encoder leads to worse performance of UAGA. The performance of UAGA is sensitive to the weight of target node classification loss $\beta$ and it is suggested to set $\beta = 0.1$. The hyper-parameter $R$ is the number of target nodes selected to construct pseudo-unknown set, and $R$=2000 achieves the best overall results.

## Conclusion

This work studies a novel O-CNNC problem, which allows target network to contain private classes unseen in the source. A novel framework named UAGA with a separate-adapt training strategy is proposed to address O-CNNC. Firstly, a rough boundary between known classes and unknown class is constructed, by training an attention-based GNN encoder and a ($K$+1)-class neighborhood-aggregation node classifier in adversarial learning. Then, negative domain adaptation coefficient is assigned to target nodes likely to belong to unknown class during adversarial domain adaptation. As a result, UAGA only aligns target nodes from known classes with the source, while pushing target nodes from unknown class far away from the source to avoid negative transfer. Extensive experiments on real-world datasets demonstrate that the proposed UAGA significantly outperforms state-of-the-art methods by a large margin in the challenging O-CNNC problem.

## Acknowledgments

This work was supported in part by National Natural Science Foundation of China (No. 62102124, No. 62362020), Hainan Provincial Natural Science Foundation of China (No. 322RC570), the Innovation Platform for "New Star of South China Sea" of Hainan Province (No. NHXXRCXM202306), the Specific Research Fund of The Innovation Platform for Academicians of Hainan Province (No. YSPTZX202410), and the Research Start-up Fund of Hainan University (No. KYQD(ZR)-22016).